\magnification=1200
\baselineskip=5mm

{\bf Proposing Quantum Relativity and Finite Program}

\bigskip

Ken'ichi Kuga\footnote{$^1$}{Department of Mathematics and Informatics, Faculty of Science, Chiba University, Chiba, Japan 263. 
 E-mail address: kuga@math.s.chiba-u.ac.jp

\noindent
Supported in part by Grant-in-Aid for Scientific Research on Priority Areas (08211211).}

\bigskip

{\bf Abstract}

We show monistic realism consisitent with quantum theory may be restored by extending the essential idea of relativity in such a way that every physical system is eligible, in principle, for an observing system. As a result, a common logical basis of quantum theory and relativity, and hence that of modern quantum gauge theories emerges. Supported by this logic, we propose to reconstruct physics solely from finite EPR complexes. Along the discussion an interpretation of String thoery is provided. Aside from conceptual appeal and a priori mathematical finiteness, our point of view drastically explains in a rather trivial fashion some basic problems which are otherwise unlikely to be resolved: Namely, the increase of entropy in macro scales, and the issue of the cosmic coincidence. In fact, the  expansion of the universe may be given a tautological reasoning in our context. 

\bigskip

1. Motivation and Program

\bigskip

There is no doubt that one of the most fundamental difficulties in modern science is the conceptual and mathematical inconsistencies between the theory of general relativity and quantum field theories.
General consensus on this problem is that the theory of general relativity, as is undoubtedly a classical theory, must be somehow cleverly quantized in a mathematically finite manner. The string theory, a modern approach to quantum gravity,however, has yet to settle the mathematical side of the problem, and more importantly, seems to be lacking in the conceptually transparent interpretation of the very basic physical principle underlying its mathematical structure.

When we look backward instead, persistent inconsisitency is already present between the theory of special relativity and Quantum Mechanics: mathematical finiteness of the basic quantities in Quantum Field Theories has not been rigorously established in reasonable generality, nor transparent physical interpretation to the calculational technique of renormalization has not been provided in a unanimously satisfactory manner.

Tracing the inconsisitency further backward, one reaches the wellknown controversy between Einstein and Bohr on the interpretation of Quantum Mechanics.

Einstein's theory of relativity, on the one hand, is based on conceptually appealing physical principles, but being a classical theory it must be modified to cover phenomena in quantum regime. On the other hand, Quantum Mechanics is experimentally perfectly verified so far, but srictly speaking, its very axioms are unescapable from somewhat misterious spiritual idealism.

The genaral idea of relativity may be phrased as clarification of invariant objective physical reality by discarding the notion of absolute observer and relativizing it: apparently different results of observations may be ascribed to a single physical phenomenon by attributing the apparent differences of observations to the relative differences between the states of the observers. Hence the very core philosophy here is (contrary to the usual idealistic relativism) nothing but the materialistic monistic realism. Hence it was very natural for Einstein to refuse the empiricist interpretation represented by Bohr and Heisenberg and others, since the essential empiricism in Physics here is ultimately of idealistic nature.

The controversy between the realist and the empiricist or idealist interpretations has not as yet been resolved in a unanimous manner. Although the realist interpretations have been unsuccesful so far, and the so called Copenhagen interpretation has been basically dominant and formally adopted, there are reasons one might feel certain uneasiness over the present status of the logical basis.

 Among others, one important reason we want to emphasize here is that, in the strict sense, when modern physicists talk about internal symmetries and gauge theories, they should be, at least implicitly, taking the side of physical realism independent of the existence of observer, and at the same time cleverly keeping away from the interpretation issue to avoid confronting the logical inconsistency. Under this circumstance, the development of modern fundamental physics has been driven inevitably by mathematical pragmatism, and as a result, one cannot provide conceptually consisitent and transparent interpretations to the core notions and mathematical derivations in many basic aspects of quantum theories beyond their mathematical utility.

\bigskip

The purpose of this paper is to show that the materialistic monistic realism consisitent with quantum theory may in fact be restored by extending the essential idea of relativity in such a way that every physical system is eligible, in principle, for an observing system. As a result, a common logical basis of quantum theory and relativity, and hence that of modern quantum gauge theories, may emerge. 

We begin by criticizing the implicit assumption of the absolute observing system in the very axioms of quantum measurement from the point of view of monistic reality. Then we present conceptual arguments leading to basic ideas composing what we mean under the name of Quantum Relativity. The point is that observer's concept such as spacetime becomes a posteriori notion and affiliated to a background physical system to which the obsever with that concept is affiliated. This point of view of conceptualization by a physical system is realy the key proposal of this paper. A main objective here is to convince the reader that this rather exotic viewpoint is in fact natural and inevitable when one faithfully sticks to the materialistic monistic realism.

We use the term observing system technically to express the combined system of the observer and the background system for the conceptualization of proper spacetime of the observer. Observer's conceptualization, then, is made possible by the existence of some symmetry of the background system, which physically corresponds to some free theory. Einstein's equivalence principle in general relativity and gauge principles may be extended here to the general equivalence of free theory of physics and geometric spacetime concept based on the same symmetry of the background system. 

If one accepts the point of view of this Quantum Relativity, it begins to make sence to reconstruct physics based solely on a finite basic structure which is not some approximation to the reality but the physical reality itself, thus restoring physical finitism. From this physical finitism, a possibility of manifestly finite reconstruction of quantum field theories arises, in the sense that mathematical divergences may be attributed to approximate mathematical conceptualization processes. Renormalizability of a theory is interpreted in this framework as a measure of applicability of this mathematical conceptualization process. For example, choice of a conceptualization corresponds to fixing of free Hamiltonian in the interaction picture, which in turn specifies the renormalization prescriptions. 

On the mathematics side, this physical finitism should shed considerable light on the deep rooted recurrnt criticism to the 'Cantor' mathematics by the mathematical finitists. The notion of the Euclidean spaces composed of the real numbers and the use of convergent series are based on the spacetime notion of classical macro-observers whose approximate conceptualization background is fairly large but still a finite physical system. In general, use of infinity is an idealization whose effective range is determined by the conceptualization basis and should be investigated. For the mathematical notion of the Euclidean spaces, the conceptualization basis is so huge and not even recognized or assumed to exist. As a result, effective range argument or some 'renormalizability' argument is missing. In this sense, 'Cantor' mathematics of the real number system is essentially of classical nature and cannot be applied directly to describe quantum phenomena. 

 As a possible mathematical basis realizing the propositions of Quantum Relativity and physical finitism, we try to begin with a set of finite graphs which may be called EPR complexes in the interpretational context. We then see how symmetries of a family of EPR complexes may be used to specify a conceptualization basis nature on it and to describe physics based on that basis. When division of conceptualization basis and observed system is universally fixed and specification of conceptualization bases is mathematically idealised and continued globally, then spacetime manifoid sturucture may emerge (strictly, in the sense of noncommutative geometry), providing a connection to the Einstein general relativity. Also from the representations of those symmetries of EPR complexes, (approximate) monoidal category structures may be obtained, which then should provide a natural connection to the rather pragmatically developed mathematical structure of conformal field theory and the integrable structure behind string 

What we must pay for this new scheme is seriously heavy. Basically we need to totally reverse the logical ordering of physics: classical mechanics is not the starting basis  of the quantizations but is an approximation to Quantum Relativistic reality, and is reconstructed by hamiltonians derived by conceptualization of macro time affiliated to macro-observing system. More specifically, Feynman path integral is replaced by a finite 'Fourier' sum and, together with macro-time conceptualization and induced causality based on entropy argument, reproduces classical Lagrangian and Hamiltonian by standard arguments in path integrals. This reversed logical ordering, however, is  the ideal direction of arguments from the point of view of monistic realism. 

The crucial aspect of macro-time conceptualization is inevitable when we thoroughly observe the materialistic monistic reality, as mentioned above. This conceptualization of macro-time point of view drastically explains in a rather trivial fashion some basic problems which are otherwise unlikely to be resolved. Firstly, we see the second law of thermodynamics, i.e., the increase of entropy in macro scales, may obtain a naturel explanation. Secondly we see, as a possible connection to phenomenology, the singularity problem of the (pre-)Big Bang models, and flatness of the global universe, and related issues,i.e.,the issue of the inflationary universe and cosmic coincidence, are explained by the high symmetry of the initial EPR complex in the sequence of macro-time conceptualization bases. In fact, the observed expansion of the universe itself may be given a possible natural (tautological) resolution in this context. 

\bigskip

This paper is organized as follows. As the primary purpose of the present paper is in the proposal of the new point of view of Quantum Relativity, the stress in the materials covered will be on their conceptual and logical consistensy in our program. 

\noindent
1. Motivation and Program

\noindent
2. Quantum Relativity

 2A) A criticism to the axioms of quantum measurement

 2B) Basic Propositions of Quantum Relativity

 2C) Conceptualization based on symmetry and Generalized Equivanence Principle

 2D) Conceptualization Basis of Macro-SpaceTime

\noindent
3. Finite Program

 3A) Finite EPR complex and Place of Renormalization

 3B) An Example: An interpretation of String Theory

 3C) Conceptualization of Time, Macro-Causality and the Increase of Entropy

\noindent
4. Connection to the General Relativity and The Cosmic Coincidence.

 4A) General Relativity as Certain Limit of Quantum Relativity

 4B) Tautological Expansion of The Universe and The Cosmic Coincidende

\bigskip

2.Quantum Relativity

\bigskip

 2A) A criticism to the axioms of quantum measurement

\bigskip

If one faithfully sticks to the materialistic monistic realism, which we do throughout, then one has to regard observers like us as part of general physical systems. As an example, imagine an obsever disintegrating into elementary particles. It is unlikely that pure physics by itself provides any natural sharp criterion which tells you the exact moment of disintegration the observer loses his status. Actually we should consider here the whole observing system including the observer instead of an isoleted observer; it already is questionable whether a single observer can observe in the usual sense when he is perfectly isolated from the rest of the universe. But even in that case, physics does not seem to provide, for example, the minimum number of particles necessary to compose an observing system.

Here we may be able to point out the incompleteness of the fundamental postulates of the Quantum Mechanics from the point of view of pure monistic realism. Namely, in the very axiom of quantum measurement, the notion of the observer or the observing system is undefined and, as a consequence, implicitly assumed as absolute and universal. This implicit absolute status of this background observing system is the essential origin of the idealism or empiricism nature of Quantum Mechanics, and responsible for various conceptual difficulties such as the problem of reduction of wave packets or the Einstein-Podolsky-Rosen issue. Temporary cure for these problems might be achieved by adding the observer to the observed system and applying the axiom of quantum measurement to this enlarged system from outside (cf. Wigner, Ref.$12$ and Machida-Namiki theory, Refs.$1, 7$). However in such arguments the absolute observing system is only pushed out of sight and is still implicitly assumed outside the physical system under c

\bigskip

 2B) Basic Propositions of Quantum Relativity

\bigskip

As mentioned above, if one sticks to the materialistic monistic realism then observing systems including observers in the usual sense must be regarded as part of general physical systems. Furthermore, there is no purely physical ground which sharply specifies observing systems among genaral physical systems. Hence, taking it the other way round, any physical system must, in principle, be eligible for an observing system. We set this as the first Proposition of Quantum Relativity.

\bigskip

Proposition A : {Theory with thorough observance of materialistic monistic realism must possess a generally applicable formalism to affiliate status of an observing system to any physical system}

\bigskip
Once one accepts this proposition, then it becomes groundless to assume our usual space-time concept or any classical physical concepts such as sharp and rigid classification of elemantary particlese as universal. In fact, the direct implication of the discovery of quantum phenomena was the denial of the continuously completed concepts of classical quantities; and the calculation of quantum field theory, in principle, need to employ all possible 'virtual' particles. Or if one considers a small system composed of unstable particles, the essentially unavoidable variation in the lifetime of the system makes it difficult to attach usual sharp time notion to this system when it is regarded as an eligible observing system. Also, the elaboreted computational technique of renormalization in Quantum Field Theories may become conceptually transparent if one accepts that the classical concepts decay in extreme scales. Therefore, we need to proceed to abandon absolute or universal status of any physical or spacetime co

\bigskip

Proposition B : {Physical and space-time concepts are affiliated to a family of physical systems which we call the conceptualization basis}

\bigskip

 2C) Conceptualization based on symmetry and Generalized Equivanence Principle

\bigskip

To clarify the essential mechanizm of observation, the distinction between the two categories of systems composing observing systems is inevitable; namely observer and conceptualization basis.

 As a physical system, an observer (with measuring devices) is usually fairly complicated and it is practically impossible to trace enormously complicated chain of interactions occurring in the process of measurement in an observer. Still worse, even a strict specification of the physical system composing the observer is usualy unavailable.
However, not all such detailed processes nor strict specifications are really essential. Note that even the complete knowledge of these processes would not determine the character of the observation by itself, due to unavoidable substantial arbitrariness in understanding what kind of quantity the observer is supposed to be measureing. Hence the character of observation is not a priori determined solely by the physical system of the observer but rather arbitrarily (and essentially whimsically) assumed by us through specification of the physical concept in question.
After conceptualization basis is specified and the revolution operator, i.e., the Hamiltonian, of the observed system with respect to macro-time is properly reconstructed, the essential mechanism of observation may be explained in a traditional language as elabolared in Machida-Namiki theory (Ref.$7$), where the practial eligibility of an observer lies in its structual complexity mathematically expressed, for example, by continuous superselection structures on the operator algebra of observables (cf.Araki, Ref.$1$).

A conceptualization basis is the family of physical systems which is responsible for an observer's formation of a physical concept, typically a geometric (external or internal) space concept.Note that our recognition of a phenomenon is ultimately based on some notion of space, and the geometric notion of space is, at least locally, equivalent to the existence of some symmetry. Then the observer can recognize and measure a physical phenomenon only when that symmetry is violated. From our point of view of materialistic monistic reality, geometric concept of space arises in an observer when the observer is saturated in a physical system with some symmetry. Hence the characteristic of a conceptualization basis is in its symmetry property. 

On the other hand, a conceptualization basis itself consists of physical systems which must be described by physics. Traditional scheme of physics, however, does not fit this purpose, for we cannot assume any a priori observer or spacetime concept from the beginning, and the spacetime concept is the essential prerequisite of the formulation of classical mechanics. 
We note, however, that recent development of conformal field theories and some other integrable models has been indicating by way of examples, that symmetry itself may essentially determine basic physical quantities. Hence the possibility of describing nature based solely on symmetries without a priori Lagrangians and Hamiltonians is actually emerging. Hence it is now recognizable that we may be able to describe the observing and observed natures of a physical system on the same footing of symmetry, realizing Principle A.  

Hence we may use symmetries in a priori different two ways in our scheme: symmetry as characterizer of a conceptualization basis and symmetry as description of physics of a general physical system. The equivalence of these two aspects of the symmetry of the conceptualization basis is nothing but the generalized version of the equivalence principle in our framework:

\bigskip

Proposition C : {Symmetry characterizes a conceptualization basis, and physics of each system in the conceptualization basis is described by the representation thoery of the same symmetry, which we call a free thoery}

\bigskip
\noindent
The somewhat vague expression of this statemant will be much clarified later in Sect.s 3A and B.

\bigskip

 2D) Conceptualization Basis of Macro-SpaceTime

\bigskip

Specification of conceptualization basis is arbitrary. This arbitrariness is similar to the arbitrarines of the choice of local coordinates around a point in Lorentzian manifold where the choice of the time axis specifies the instantaneous motion of the observer in Einstein's relativity. One fundamental difference here, however, is that in our framework, conceptualization bases are usually not arranged along any particular geometric space (manifold) expressing classical locality and causality. As we abandoned the privileged status of the macro spacetime, conceptualization bases may well be intertwined or nested with no regard to the arrangement with respect to the macro spacetime.(cf. Sect.4A)

 If the concept of macro-spacetime looks comparatively universal througout the universe, at least among sufficiently large observers, then we need to attribute that particular characteristic to that of its conceptualization basis. 
A natural assumption seems to be the following:

\bigskip

Proposition D (Hypothesis on the macro-spacetime conceptualization basis) : {The conceptulalization basis of the macro-spacetime consists of the systems of mass-zero particles interrelated in the eary universe.}

\bigskip

3. Finite Program

\bigskip

Although strict finitism for Mathematics may have been repeatedly demonstrated to be unpractical, finitism for physical monistic reality seems conceptually appealing. In fact Quantum Relativistic viewpoint provides the required conceptual support to this: As we are led to abandon the fundamental status of continuous notion of spacetime in the course of the above consideration of quantum relativity, it is rather natural for us to base physics on some finite structure. Aside from the conceptual appeal, it has the mathematical merit of making the theory a priori consistent and finite.

Here we want to emphasize again that mathematical notion of the continuously completed real numbers is an artificial idealization of the macro spacetime, formed essentially through our classical macro-experiences. In this respect, it is no accident that the invention of the Newtonian mechanics and the mathematical invention of calculus took place simultaneously in the history of science. The real number completion was needed to smoothly mathematize calculus. 

One of the first modern approaches in this direction may be the Regge analysis (Ref.$10$) where Riemannian manifoids are replaced by $4-$dimensional simplicial complexes. In simillar spirits, given a quantum field theory, one may set the cut-off energy near the highest possible energy level for the given theory to be practically meaningful. Then that cut-off energy determines the minimal spacing of the lattice approximation and removes divergences in the obvious manner. 
A conceptual difficulty in these approaches is that the naive uniform lattice approximation itself can hardly be seem to represent some finite physical reality itself. As a result, stability of the theory with respect to the spacing, i.e., renormalizability, is essentially required, whose mathematical finiteness is usually hard to establish. In fact, the choice of such cut-off energy becomes a very delicate issue when the theory does not assume a continuous limit. As an exapmle one may take the $4-$dimensional ${\varphi}^4$ theory. The cut-off energy may then easily been seen to severely restrict the coupling constants at low energy levels. Hence it is desirable if such cut-off energy level, or equivalently the minimal spacing is given some intrinsic status which is in principle autonomously determined.

Hence we propose to start with some basic finite structure representing some monistic physical reality from the beginning, and formulate the observing systems in this finite structure in such a way that any subsystem is in principle elligible for an observing system. The spacetime notion of an observing system is formed by its conceptualization basis. The cut-off energy and other approximation parameters should come from this finite conceptualization basis. In this flexible framework, many modern theory may be interpreted essentially though the automorphism group determined by the symmetry structure of the conceptualization basis. Then, in principle, that the traditional quantum field theroy and then classical mechanics will be reproducable in our scheme.

An important aspect of this scheme is the need of macro-time conceptualization process. Since we abandoned the absolute status of spacetime, we need to define it solely from the basic finite structure. We do so using the ordering of observing systems according to the decay of the conceptualization bases of the macro-space time which we assume are the huge systems composed of massless particles interrelated in the early universe (Proposition D). This definition of macro-time makes the second law of thermodynamics almost a trivial statement and reproduces the causality notion on which the whole development of physics has been based. Then the reproduction of classical formalism and quantization procedures follow just by reversing the logic of the path-integral method. It also explains autonomous expansion of the universe and the observed "cosmic coincidences" which are otherwize unlikely to be understood. 

\bigskip

 3A) Finite EPR complex and Place of Renormalization

\bigskip
The paradoxical nature of the Einstein-Podolsky-Rosen issue (Refs.$2, 4$) is resolved once we abandon the absolute status of macro-spacetime.
Then the Einstein-Podolsky-Rosen paradox is interpreted as the claim of priority in substantiality of interrelated particles over the notion of macro spacetime in our framework. This priority of 'EPR connections' over the classical spacetime continuation is compatible with the physical finitism mentioned above and motivates us to introduce the finite universe ${\Upsilon}_{\Phi}$ as the basic structure below. 

Our naive formalism in this subsection will be in no way intended to set any rigid mathematical basis to the program. Rather the main purpose here is to see the potentiality of reconstructing usual physics beginning solely with some finite structure. We demonstrate this by justifying and interpreting the various existing theories including calculational technique of renormalization procedures and the modern integrable string theories by finding their proper places in our framework. 

Therefore at this very basic and general stage, it seems inessential which mathematical concept to use to represent the raw finite physical reality. The essential point here is that we assume some basic finite structure which is safe from 'Cantor' mathematics.

So let us begin with just a set of finitely many objects:
${\Phi}=\{o_1, o_2, \dots, o_N\}.$

A simplest structure on it may be provided by a specification of a family of subsets of $\Phi.$ Essentially equivalently one may consider abstract simplicial complex structures with vertices in ${\Phi}.$ 
We further restrict ourselves to $1-$dimensional complexes, i.e., graphs, in this paper. Hence we set

$ 
{\Upsilon} =
{\Upsilon}_{\Phi}=
\{E;\ abstract\ simplicial\ 1-dimensional\ complexes\ with\ obj(E) \subset {\Phi}\}
$

\noindent
and call it the universe based on the finite set $\Phi.$
We call each element $E$ of $\Upsilon$ an {\it EPR complex} in the universe ${\Upsilon},$ and the edges of an EPR complex $E$ {\it EPR connections} of $E.$
For $E, F \in {\Upsilon},$ we write $E \le F$ when $obj(E) \subset obj(F)$ and two objects of $E$ are joined by an edge whenever they are joined in $F,$ i.e., when $E$ is a fullsubgraph of $F.$ Then $({\Upsilon}, {\le})$ becomes a partially ordered set. We call a maximal element of $({\Upsilon}, {\le})$ an {\it aspect} of ${\Upsilon},$ and denote by ${\cal A}{\Upsilon}$ the set of all aspects of ${\Upsilon}$ : ${\cal A}{\Upsilon} = \{E \in {\Upsilon} | obj(E) = {\Phi} \}.$ 
Aspects are the fundamental independent states of our universe ${\Upsilon}_{\Phi}$ providing, in principle, orthonormal states, i.e., the inner product for the Hilbset space in each theory. We need to restrict ${\cal A}{\Upsilon}$ to a smaller subset for each individual theory: When an EPR complex $E\in \Upsilon$ or some class of them is fixed, uaually as the conceptualization basis and macro-observers, we may consider 
${\cal A}(E) = \{ A \in {\cal A}{\Upsilon} | E \le A \}.$  
In practice we further restrict ${\cal A}(E)$ to a much smaller subset to form the orthonormal basis of the Hilbert space of the thoery, assuming most contributions from ${\cal A}(E)$ cancel out and are inessential. An explicit example will be given in Sect. 3B. Aside from the role of the genarator of Hilbert space bases, a sequence of aspects is necessary to generate sequence of macro-time conceptualization bases which will be explained in Sect. 3C.  

When an aspect $A \in A{\Upsilon}$ is fixed and a set of EPR complexes smaller than $A$, say $E_1, E_2, \dots ,E_s \le A$, is given, we can form a larger EPR complex ${*}_A{\{E_i\}}_{i=1}^s = E_1 {*} E_2 {*} \dots {*} E_s \in \Upsilon,$ the {\it join} in $A,$ which is defined to be ${\min}\{ F \le A | E_i \le F, i=1, \dots ,s \},$ i.e., the smallest EPR complex of $A$ larger than or equal to any of $E_i, i=1,\dots ,s.$  This provides a typical way to extend conceptualization basis to reach free (integrable) theory (c.f. Sect.3B).
\bigskip

The issue of Physics in this setting is to describe how the universe ${\Upsilon}$ looks from various observing systems $E$ {\it inside} of $\Upsilon$. For this purpose each element of ${\Upsilon}$ need to play the dual roles of observing and observed systems. As proposed in Section 2C, we want to base this double-fold function on specifications of symmetries to elements of ${\Upsilon}.$ 

Hence the primary mathematical object might be the simplicial automorphism group ${\Sigma}_E = Aut(E)$ for each $E\in \Upsilon$. However, this simple universe ${\Upsilon}_{\Phi}$ is not really simple at all in this respect when the cardinality of $\Phi$ is not small. In fact, for example, it is not hard to see that any finite group $G$ can be realized as the symmetry group $G = {\Sigma}_E$ of some $E\in {\Upsilon}_{\Phi}$ for large enough $\Phi$.

 Hence to bridge the finite $\Upsilon$ and the traditional formulations of Physics, we need to employ some approximate methods. For example we may say an EPR complex $E$ has an approximate symmetry expressed by, say a Lie group $G$, when there is a quotient group $G/N$ and a 'reguralization' ${\Sigma}_E\to G/N$ is provided. Typically, such an $E$ is approximately a homogeneous space $G/H$ of $G$.

 Aside from the technical difficulty in handling large finite symmetries, this compromise is somewhat inevitable in practice, since we are also inside of $\Upsilon$ and not a priori given the details of the EPR complex $E$ and there is usually no way for us to specify $E$ exactly in advance. 
As a result, contrary to the logically ideal direction, it may well happen that we can only start by assuming some symmetry $\Sigma$ first and only specify a class ${\cal E} = \{E_t|t\in T\} \subset {\Upsilon}$ of EPR complexes with an approximate symmetry expressed by $\Sigma$.
 The relevant symmetry $Sigma$ then is not only an approximation of the automorphism of each $E_t$ but rather the symmetry among all the $E_t$'s. 
 (The essential arbitrariness in the choice of $\Sigma$ has already been discussed in Sect.2C.) We call such an ${\cal E}$ a conceptualization basis with symmetry ${\Sigma}.$
Such a conceptualization basis may well composed of practically nearly infinite number of objects and as discrete groups are not always mathematically easily handled, we may frequently take $G$ to be a Lie group or a Lie algebra in some infinitesimal operator formalism. 

\smallskip

Clarification of the precise meaning of the 'reguralization' ${\Sigma}_E\to G/N,$ or typically, the approximation of the finite conceptualization basis $E$ by a possiblly continuous homogeneous space $G/H,$ becomes a central issue. This is where the whole regularization and renormalization technology, especially the renormalization group mehtod, finds its place in our scheme. For example, the finite conceptualization basis in momentum space or more generally in phase space gives natural interpretation to the phase-spece expansion in, say, Pauli-Villars regularization, from which renormalization algorithm based on the so-called forest formulas emerges. Very roughly, the renormalizability should correspond to the approximability of an a priori unknown finite conceptualization basis by an artificially specified continuous symmetry model.(cf. e.g.Ref.$11$)
To keep simplicity at this very basic stage, however, systematic reinterpretation and reorganization of renormaliation theory in this point of view, and hence further clarification of the approximation issue, cannot be dealt with in the present paper and will be postponed to later papers. We here therefore somewhat 
crudely use some infinite contunuous groups without further comment. 

\smallskip

Typically a class of EPR complexes composing a conceptualization basis ${\cal E}$ arises when there is a big EPR complex $E\in \Upsilon$ with large symmetry $\Sigma$, and a family of aspects ${\cal A} = \{A_t| t\in T\} \subset {\cal A}{\Upsilon}.$ Then one obtains ${\cal E} = \{ E \wedge A_t | t\in T\},$ where $E \wedge A_t = {\inf}\{ E, A_t \},$ i.e., the largest EPR complex $I \in \Upsilon$ with $I\le E, I\le A_t.$
Each (instataneous) observer $O_t \in \Upsilon$ is in one of these aspects, say $O_t \le A_t,$ and for this observer the observation will be based on the EPR complex $E_t$ with symmetry possibly broken down to a smaller one from $\Sigma.$ 
Hence for a class of the observers $\{O_t\},$ the recovering of the original symmetry $\Sigma$ is inevitably a trial and error process, which is practically the case for the modern development of fundamental physics.

Suppose further we have the simplest situation where each EPR complex $E_t = E \wedge A_t$ is essentially the same with the aspect $A_t.$ By this we mean all the objects in $A_t$ which are not in $E_t$ are EPR-isolated. As observers and observed systems are in the aspects, there is no practical observer in this setting and observed systems are contained in the conceptualization basis. This is the idealized free situation specified in Proposition C. Hence, aside from the issue of the approximability by $\Sigma,$ the symmetry $\Sigma$ not only characterizes the conceptualization basis but also describes the physics.  Specifically the representation theory of $\Sigma$ defines the concept of free particles and a vector in one-particle Hilbert space is a linear combination of them, which we call a Fourier sum. The mathematical construction of the Fock space of those free particles comes in here naturally to idealize the differences among the EPR complexes in ${\cal E}.$ The large common portions are regarded as 
Expressing vectors in free fermionic Fock space by formal semi-infinite differential forms, for example, is a way to take this common portion into consideration in the simplest manner. Creation and annihilation operators are interpreted here as the infinitesimal generators of the deformations of an EPR complex ${E_t}$ to another EPR complex $E_{t'}$ in ${\cal E}$. 

The most basic example is the macro-spacetime conceptualization basis ${\cal E}_M = \{ E_M(t) \}_{t\in T} \subset {\Upsilon}.$ We hypothesized in Proposition D that each $E_M(t)$ is composed of massless particles interrelated in the early stage of the universe. We shall discuss the interpretation of the parameter $t \in T$ as the macrotime parameter in the next subsection.  The symmetry $\Sigma$ may then taken to be the Poincar{\'e} group and $H$ the Lorentz group yielding Minkowskii space ${\Sigma}/H$ approximating the energe-momentum space of finitely many "free" particles in $E_M(t).$
The finiteness of $E_M(t)$ gives autonomous cutoff in the momentum space, removing divergent integrals in an obvious manner.

In general situation the aspects are larger than the EPR complexes in the conceptualization basis, and further more there can be more than one distinct conceptualization basis. Simple Fourier sum will be replaced by multi-stage sum which should formally yield path integrals.
Rigorous direction to solve a theory is to extend the EPR complexes in the conceptualization basis to a larger complexes forming a conceptualization basis where the theory becomes free and solble in principle.

A typical way to extend EPR complexes may be found in the Borel-Weil type construction (cf. Ref.$9$). That is, we can construct (irreducible) representation spaces as holomorphic sections of some complex line bundles over the homogeneous spaces like $G/(max torus) \equiv G_{\bf C}/(Borel),$ where we assumed that the symmetry here is a compact group ${\Sigma}=G$ and $G_{\bf C}$ is a complexification of $G$ and $(Borel)$ refers to a standard Borel subgroup of $G_{\bf C}$.
Such line bundles and sections are constructed though the EPR-connections from the conceptualization basis $E \approx G_{\bf C}/(Borel)$ to another set of objects forming a basis of a complexs vector space $V$. In fact, considering the EPR-connections as defining a map from conceptualizationbasis $E$ to the vector space $V$ one can pull-back the rays of $V$ to a line bundle on $E$ which supports the  representation of $G$. 
 A standard approach to comformal field theory is formulated in this line of argument where the group is actually the virasoro group and homogeneous space is a certain dressed moduli space of Riemann surfaces. Well behaved sections, the tau functions, play the basic role of the physical vacua 
Hence the technology of conformal field theory may be regarded as that of computing 'Fourier coefficients' comming from the EPR-connections and hence obtaining correlation functions and basic physical quantities. We will discuss the interpretation of the string theory in this line of argument in the next subsection.(cf. e.g. Refs.$6, 13$) 

\bigskip

 3B) An Example: An interpretation of String Theory

\bigskip

To see the above statement more concretely in our quantum relativistic framework, assume we have an 'atlas' $\{({\cal E}_i,{\Sigma}_i)\}_{i=1}^n$ of $\Upsilon$ composed of conceptualization bases ${\cal E}_i \in {\Upsilon}$ with symetry ${\Sigma}_i, i=1, \dots ,n.$ We assume each ${\Sigma}_i$ is given as a Lie group $G_i$ and each $E \in {\cal E}_i$ is approximated by the Lie group $G_i$ or by a homogeneous space of it.
We further asumme that one of which, say ${\cal E}_1,$ is the macro-spacetime conceptualization basis ${\cal E}_M$ mentioned above.
The simplest Lie group is the circle $S^1$ which we assume to be the case for $G_2 = S^1.$ 
Although we usually choose the macro-spacetime conceptualization basis to form the observing system for the theory, each of these conceptualization basis provides in principle an eligible observing system. So one can choose, for example, ${\cal E}_2$ as the basis of the theory. 
This means one expresses objects in other conceptualization bases or more generally in ${\Upsilon}_{\Phi}$ in terms of the EPR-connections in the objects in ${\cal E}_2$, which is, in practice, by Fourier expanding them in terms of the representation theory of $G_2 = S^1.$ 
Then the loop groups (algebras) arise naturally as the function spaces from $G_2$ to $G_j$ idealizing EPR-connections from $S^1 \approx E \in {\cal E}_2$ to $G_j \approx F \in {\cal E}_j : Map(S^1,G_j) \approx E*F,$ where some ambient aspect $A$ is fixed, presumably $A = {\bigcup}_tA_t \in {\cal A}{\Upsilon},$ the union of aspects parametrized by macro-time parameter (cf. Sect. 3C).  Kac-Moody symmetries arise as (central extension of) the symmetries of the combined complexes $E*F.$  The (appropriately regularized) Casimir operator,i.e., the Sugawara construction, affiliates Virasoro aligebra. After interpreting it as energy-momentum tensors, one may obtain conformal field theory of genus zero. 

Now suppose that more than one $G_i$'s, say from $G_2$ to $G_s, (s < n),$ are the circles $S^1,$ and typical EPR complexes $E_2 \in {\cal E}_2, \dots ,E_s \in {\cal E}_s$ are interrelated in an aspect, i.e., $E_2*{\dots}*E_s$ is not the simplest disjoint union. For a geometric image, one may consider a triangulated Riemann surface whose link complex at each point is one of $E_i, 2 \le i \le s$ approximated by the circle $G_i = S^1.$ This setting is 'conformally invariant' in the sense that the link circles $E_i$'s are rigid in the rotational direction but the metric itself on the surface is undefined and irrelevant. Hence idealizing each $E_2*{\dots}*E_s$ as a conformal structure on a Riemann surface of specific genus, the totality ${\cal E}_2*{\dots}*{\cal E}_s$ is approximated by the union of Riemann moduli spaces of all genera. 

Dressed moduli spaces, which themselves are a key construction to embed them into a certain infinite Grassmannian, may be regarded as an extended join $E_2*{\dots}*E_s*F$ for some $F \in {\cal E}_p, s+1 \le p \le n ,$ i.e., the EPR connections between each $E_i, 2 \le i \le s,$ and $ F.$ The integrable structure behind string theory is in fact caputured in this line of argument (cf. Ref.$6$).
Note that EPR connections are used in two different ways here: they are used to define mappings from an EPR complex regarded as a conceptualization basis to another EPR complex; they are also used to form a larger EPR complex of higher symmetry structure. Such an enlarged EPR complex characterized by high symmetry makes the theory free and integrable. This is an example of the general equivalence expressed in Proposition C.

The essential input required to interpret some conformal field theorys formulated on Riemann surfaces of arbitrary genera as a string theory is the identification of the macro-spacetime directions, i.e., identification of the energy-momentum tensor. Such a macro-spacetime direction is provided by the EPR connections between $E_2*{\dots}*E_s*F_p*{\dots}*F_q$'s described above and the macro-spacetime conceptualization basis ${\cal E}_1 = {\cal E}_M = \{E_M(t)\}_{t\in T}.$  The resulting maps $Riemann\ Surfaces \to {\bf R}^{1,3}$ also justifies somewhat peculiar formulation of the string theory based on the $2-$dinensional world sheet. 

Hence, in short, the string theory is essentially the first nontrivial example based on non-macro conceptualization basis, and integrability is in the high symmetry structure of extended EPR complex. The only essential input required is the provision of the macrotime direction by the macro conceptualization basis.

 We will see how the "arrow of macro-time" may be defined inside the universe $\Upsilon$, and hence classical formulation of physics arises, in the next subsection.

\bigskip

 3C) Conceptualization of Time, Macro-Causality and the Increase of Entropy

\bigskip

Classical mechanics is formulated in such a way that it describes the time evolution of a physical system. This formalism is deeply rooted on the very recognition process of nature through (macro-)causality. However, since we have abandoned such an absolute status of the concept of macro-time as it is affiliated to macro-observers like us, we need to begin with the problem of this macro-time conceptualization in order to be able to reconstruct classical mechanics and traditional quantization procedures in our scheme. This means, in particular, we abandon the very notion of causality as a basic strict notion: approximate macro-causality should reappear only after we conceptualize the macro-time. This abandonment of the fundamental status of locality and causality seems inevitable ultimately in the end, since for example the PCT theorem based on these fundamental prerequisites is incompatible with the second law of thermodynamics which is universally valid in macro-scale. 

Recall we set the following hypothesis on the conceptualizaton basis of macro-time (Proposition D): {The conceptulalization basis of the macro-spacetime consists of the systems of mass-zero particles interrelated in the eary universe.} 

In reality the particles interrelated in the early universe is decreasing as they are constantly caught by stars and various materials in the universe. Hence there are a huge number of aspects $A(t) \in {\cal A}{\Upsilon}, t \in T$ separating instaneous observers $O(t) \le A(t).$ Each observer $O(t)$ conceptualizes macro-spacetime based on the EPR complex $E_M(t) = E_M \wedge A(t),$ where $E_M$ is the original macro-spacetime conceptualization basis. Now it is the problem of multiuniverse point of view whether we can reasonably linearly order these conceptualization bases or not. Instead of going into this issue, we here only want to claim here that from the macroscopic asymmetry between the past and the future the natural way to (quite possibly non-linearly) order them is to order them accorrding to some entropy notion. 
Hence to conceptualize macrotime, it is in fact sufficient to assume that we have a chain of aspects $\{ A(t_i) \}_{i=I}^J\subset \{A(t)\}_{t\in T}$ with conceptualization bases $\{ E_M(t_i) \}_{i=I}^J$ in the sense that $E_M(t_i)\supset E_M(t_j),$ simplicial subcomplex, for $i<j.$  As our universe $\Upsilon $ is at least a partially ordered set with respect to the simplicial inclusion relation $\supset$, and we simply took a totally ordered subset. The reparametrization of the parameter $\{t_i\}_{i=I}^J$ by a monotone increasing sequence of actual time parameters, say  $\{ {\tau}_i\}\subset {\bf R}$ will depend on the choice of the observers $O(t_i))$, as the actual time parametar has no invariant meaning. The ordering, however, is fixed. This is our definition of the "arrow" of macro-time.
Hence we say a system, i.e., an EPR complex $F \in \Upsilon$ exists at time ${\tau}_i$ with respect to a fixed macro-time parameter $\{{\tau}_i\}$, when $F \le A(t_i).$  

\medskip

The second law of thermodynamics may be given a natural explanation from our definition of the "arrow" of macro-time. 

To understand it, note first that one can rather easily 'prove' an increase of entropy type theorem under certain conditions, e.g., Markov type H theorems. However, as the time evolutions of an isolated system classically form a group, what one actually proved turns out to be an entropy invariance theorem with respect to the time evolution. Hence the essential problem in this issue is how one can attain irreversibility for an seemingly isolated system. From our point of view, a classically isolated system $S\in \Upsilon$ is actually connected to other parts of the universe to form an aspect $ A(t_i)$ containing, in particular, an element $E_M(t_i)$ of a fixed chain of macro-time conceptualization $\{E_M(t_h)\}_{h=I}^J.$ 
Most of such connections among $A(t_i)$ do not have any classical dynamical influence on the isolated system $S,$ but changes the number of possible states of $S.$ Then the irreversibility of the direction of the chain $\{E_M(t_h)\}_{h=I}^J$ explains the irreversibility of the passage of macro-time. Specifically, as time passes, $E_M(t_i)$ decays by definition and leaves smaller $E_M(t_j) \subset E_M(t_i).$ Then the obsevables of time and position of the observer connected to $E_M(t_j)$ have less resolution power than that of the observer connected to $E_M(t_i),$ sinse time and position conceptualise though (finite) Fourier expansion based on $E_M(t_h),\ h=i, j.$ Then it can be seen that the properly defined entropy of such observables actually increases. (cf.e.g. Ref.$8$).

\bigskip

Classical causality follows along the same line of argument. Here again note first that the quantum mechanical expression of special relativistic causality, i.e., the vanishing of the commutators of spacelike separated fields follows just by manipulating the commutator of the Fourier expanded fields (cf. e.g. Ref.$5$). By classical causality we mean our experience that the cause of a result must occur in advance.From our point of view, an object has connections both from the past and future, since such division does not have any invariant meaning to the object under observation. Hence we add up all the contribution from past and future to form finite Fourier summation. However the contribution from the future is negregible compared with that from the past in macro-scale, since contributions from objectds connected only to smaller conceptualization basis $E_M(t_j)$ oscillate randomly.        
Here we want to add that the classical causality has been regarded as something most fundamental in our recognition process and formalism of classical mechanics. From the point of view of our monistic reality, however, this should be interpreted to be the 'philosophical arrow of time' in the terminology of Hawkings, which refers to the usual experience that we remember tha past but we can never 'remember' the future. This experience is only a concequence of the above classical causality.

\bigskip
 4. Connection to the General Relativity and The Cosmic Coincidence.

\bigskip

 4A) General Relativity as Certain Limit of Quantum Relativity

Einstein's formulation of the general relativity is decisively influenced by the mathematical notions of (pseudo-)Riemannian manifolds and Ricci and Levi-Civita's 'absolute differential calculus'. The basic idea here is the classical locality. The equivalence principle formulated in the general covariance almost inevitably leads to the action integral composed of the curvature tensor. Variational principle then routinely produces the Einstein field equation (in vacua). Hence to see our Quantum Relativistic setting approaches General Relativity in a certain idealized limit, it essentially suffices to see the parallelism between them on the principle level.    

Classical locality is supported by the mathematical notion of locally euclidean topological spaces underlying spacetime manifolds for General relativity or some fiber bundles over them mathematizing internal gauge theories. 
In Quantum Relativity, the finite EPR complexes $\Upsilon$ are the underlying structure supporting locality.  That is, the first priority is given to the EPR connections over the classical infinitesimal continuation of spacetime.  

In a spacetime manifold, essentially arbitrary choice of local charts, i.e., specification of coordinate systems on open subsets, specifies the (instantaneous) observing system moving along its time axis. It is assumed that one can always find some coordinate system around each point of the manifold so that gravitation-free special relativity holds infinitesimally near that point in terms of the coordinate system. This assumption of course is based on the Equivalence Principle. Specification of these infinitesimally free local observers' orthonormal coordinate systems is nothing but the metric tensor on the manifold. Then relation between different observers is given by the coordinate transformations obtained via the underlying topological space. 
  To this corresponds an essentially arbitrary choice of symmetry model based on EPR complexes of a conceptualization basis with some symmetry $\Sigma.$ This symmetry not only characterizes the conceptualization basis but also describes physics of free theory (Proposition C). This generalization of the Equivalence Principle is a reflection of the modern gauge theories generalizing Uchiyama's (external) gauge principle and Yang-Mills (internal) gauge principle. 
Hence, the only essential difference in this parallelism is the configulation of 'local charts':In a spacetime manifold, local charts are assumed to be arranged along the mathematical notion of a manifold which is a mathematical creation based on the privileged emphasis on idealized classical locality. In our framework, however, the conceptualization bases are usually not arranged along any particular geometric space (manifold) expressing classical locality and causality. Rather, we assumed a nested configuration of spacetime conceptualization bases with the maximal EPR complexes, in order to explain universal conceptualization of spacetime concept.(Sect.2D) 
Hence, when this nested configuration is replaced by a configuration with evenly distributed small spacetime conceptualization bases, classical locality reappears and the general covariance based on this classical locality reproduces the Einstein field equation.

\bigskip

 4B) Tautological Expansion of The Universe and The Cosmic Coincidende

In (pseudo-)Riemannian geometry the fundamental quantity is the metric tensor. From our point of view, however, the distance between two points is defined after they are expressed as finite Fourier sums on a macro-conceptualization basis. Hence, on the level of Fourier modes, the distance is the average difference in the phase of their modes.
This definition of the classical distance by the finite Fourier series and attribution of the passage of macro-time to the decay of the macro-spacetime conceptualization basis tautologially implies the increse of the distance between two typical points in the universe. In fact if the number of the finite Fourier basis elements decreases then the sharply localized original universe begins to diffuse. Objects detached from the macro-spacetime conceptualization basis begin to form materials and eminate local-spacetime conceptualization bases causing local variations of the original macro-spacetime.  Here the finiteness assumpution is essential, since decrease of finitely many Fourier basis elements narrows their limits of extension. Otherwise dilution of infinitely many Fourier basis elementa would only cause the dilution of the functions without changing thier dintances.

Note that in this picture, the classical limit in subsection $4A$ is an unreal idealization. Suppose the distance between two points $P,Q$ in the universe are extremely large. Then the only possible chains of EPR-connections between the two objects at $P$ and $Q$ necessarily go back to an almost maximal macro-spacetime conceptualization basis, which is flat by definition.

The actual evolution of the real universe is a fairly complicated process consisting of layers of different eras. Hence it is unlikely to think we today are directly connected to the very beginning of the universe. That is, the macro-spacetime conceptualization bases for us usually do not extend to the very biginning basis. It is, however, only necessary to assume the flatness of the earliest stages of the universe. In fact it is an implication of theories on the inflationary universe, that if the universe was highly symmetric and flat before the Big Bang, then the traditional physics explains the current state of the universe, including the issue of "Cosmic coincidence."

\bigskip

{\bf References}

\bigskip
\noindent
$1.$\  Araki, H.: A remark on Machida-Namiki theory of Measurement, Prog. Theor. Phys. 64, 719-730 (1980) 

\smallskip
\noindent
$2.$\  Bell, J. S., Physics, 1, 195 (1964) 

\smallskip
\noindent
$3.$\  Christ, N. H., Friedberg, R., and Lee, T. D., Nucl. Phys. B202, 89 (1982)

\smallskip
\noindent
$4.$\  Einstein, A., Podolsky, B., and Rosen, N., Phys. Rev., 47, 777 (1935)

\smallskip
\noindent
$5.$\  Itzykson, C. and Zuber, J.- B.: Quantum Field Theory, McGraw-Hill, (1980)
\smallskip
\noindent
$6.$\   Kawamoto, N., Namikawa, Y., Tuchiya, A., and Yamada, Y.: Geometric realization of conformal field theory on Riemann surfaces, Commun. Math. Phys. 116, 247-308 (1988)

\smallskip
\noindent
$7.$\   Machida, S., and Namiki, M.: Theory of measurement in quantum mechanics, Prog. Theor. Phys. 63, 1457-1473, and 1833-1847, (1980)

\smallskip
\noindent
$8.$\   Nakamura, M., and Umegaki, H,: On von Neumann theory of measurements in quantum statistics, Math. Japan  7, 151-157 (1962)

\smallskip
\noindent
$9.$\   Pressley, A., and Segal, G.: Loop groups. Oxford University Press, 1986

\smallskip
\noindent
$10.$\  Regge, T., Nuovo Cim. 19 558 (1961)

\smallskip
\noindent
$11.$\  Rivasseau, V.: From perturbative to constructive renormalization, Princeton Univ. Press, (1991)

\smallskip
\noindent
$12.$\  Wigner, E. P., Am. J. Phys. 31 (1963), 6.

\smallskip
\noindent
$13.$\  Witten, E.: Quantum field theory, Grassmannians, and algebraic curves. Commun. Math. Phys. 113, 529-600 (1988)

\end